\input mates.sty
\input epsf.sty
\def\hb{\hfil\break}

\def\registrado{{$\scriptstyle\bigcirc$}\kern-.7em ${\scriptscriptstyle R}$}
\def\plaintex{\setbox1=\vbox{\hsize80mm \vskip.1cm
 Typeset with Plain\TeX\kern.5em \& fjy.maq'ros\kern-.16em
\setbox0=\hbox{\raise1.1ex\hbox{\registrado}}\copy0   
\vskip.1cm}
\setbox2=\vbox{\special{picture star scaled 600}}
\line{\box1\kern2em\box2}}
\rightline{FTUAM 96-12 }
\rightline{March, 1996}
\vskip.5cm
\centerline{{\bf THEORY OF SMALL }$x$ {\bf INCLUSIVE PHOTON SCATTERING, I}}
\vskip1cm
\centerline{{\bf  F. J. Yndur\'ain}\footnote*{e-mail: fjy @ delta.ft.uam.es}}
\vskip.3cm
\centerline{\sl Departamento de F\'\i sica Te\'orica, C-XI}
\centerline{\sl Universidad Aut\'onoma de Madrid}
\centerline{Canto Blanco, 28049-Madrid}
\vskip1.5cm
{\bf ABSTRACT}.-
\vskip.3cm
\setbox0\vbox{\hsize120mm In the early eighties, L\'opez, Gonz\'alez-Arroyo
 and the present author proved that, 
if at a given $Q_0^2$ large enough for perturbative QCD to be valid,
 structure functions behave as a power of $x$ for $x\rightarrow 0$, 
then for all larger $Q^2$ one has
$$F_2(x,Q^2)\simeq B_S[\alpha_s(Q^2)]^{-d_+}x^{-\lambda}
+B_{NS}[\alpha_s(Q^2)]^{-D_{11}}x^{0.5},$$
$$F_G(x,Q^2)\simeq B_G[\alpha_s(Q^2)]^{-d_+}x^{-\lambda}$$
$$R(x,Q^2)=\frac{r_0\alpha_s(Q^2)}{\pi},$$
with $D_{11},\,d_+,\,B_G,\,r_0$ calculable in terms of $B_S,\,\lambda$. Moreover, it was 
suggested that the ``hard" part of the scattering cross section for 
real photons (Compton scattering) obeys a similar law, so that
$$\sigma_{\gamma p}\simeq B_{\gamma p}s^{\lambda}+A_{\gamma p}\hat{\sigma}^P,$$
with a value of $\lambda$ comparable to that in the expression for the structure functions,
 and where $\hat{\sigma}^P\sim\log^2s$ 
is a universal, Pomeron-type cross section, and $A_{\gamma p},\, B_{\gamma p}$ 
are constants. In the present 
paper it is shown that the recent HERA measurements may be described by these 
formulas, with 
a chi-sqared/d.o.f. substantially less than unity, and with values 
of the parameters compatible with those of the old fits of the '80s. Moreover,
 further discussions are presented both on the low $Q^2$ limit, and the transition 
between Compton and deep inelastic scattering, in particular 
in connection with possible saturation of the coupling constant 
$\alpha_s(Q^2)$ at small $Q^2$;  and on the ultra high energy limit, and how one 
might test the so-called BFKL conjecture,
$$\lim_{x\rightarrow 0\atop Q^2\rightarrow \infty}F_2(x,Q^2)\sim x^{-c_0\alpha_s}.$$\hb
With respect to the last we find some 
evidence against it, at least at the HERA energies.}
\centerline{\box0}
\vskip0.2cm

\vfill\eject
\noindent 1.-{\bf INTRODUCTION}
\vskip1cm
We consider in this note high energy inclusive scattering of (virtual or real) photons 
off protons:
$$\gamma^*(q)+p(p)\rightarrow {\rm all}.\eqno (1.1)$$
In the case of {\it real} photons we will call the process Compton scattering; 
for {\it virtual} photons we have DIS (=deep inelastic scattering). In this 
last situation we consider the region of very small $x$ with\footnote*{More 
details on notation may be found in ref.1}
$$x=Q^2/\nu,\;\nu=p\cdot(p+q),\;Q^2=-q^2.\eqno (1.2)$$
Compton scattering may be thought of as the limit of DIS when
 $Q^2\rightarrow 0,\,x\rightarrow 0$ in such a way that
$$Q^2/x\simeq s,\;s=(p+q)^2.\eqno (1.3)$$
To a large extent the present work may be considered as an {\it aggiornamento} of 
the old analysis of C. L\'opez, A. Gonz\'alez-Arroyo and the present
 author, applying it to the recent HERA DIS and Compton data. In this respect it is shown 
that the new data are fitted very well by the formulas derived from 
QCD in the small $x$ region, and that the expressions
$$F_2(x,Q^2)\simeq B_S[\alpha_s(Q^2)]^{-d_+}x^{-\lambda}
+B_{NS}[\alpha_s(Q^2)]^{-D_{11}}x^{0.5},$$
$$F_G(x,Q^2)\simeq B_G[\alpha_s(Q^2)]^{-d_+}x^{-\lambda}\eqno (1.4)$$
$$R(x,Q^2)=\frac{r_0\alpha_s(Q^2)}{\pi},$$
with $B_G,\,D_{11},\,d_+,\,r_0$ calculable in terms of $B_S,\,\lambda$, give an 
excellent approximation to the data for $x\lsim 10^{-2}$ (Sect.2). Likewise (Sect.3) 
we find that the very high energy Compton cross section is still 
 correctly described by 
the formula
$$\sigma_{\gamma p}\simeq B_{\gamma p}s^{\lambda}+A_{\gamma p}\hat{\sigma}^P,\eqno (1.5)$$
with a value of $\lambda$ similar to the $\lambda$  in (1.4), and where $\hat{\sigma}^P\sim\log^2s$ 
is a universal, Pomeron-type cross section.
 $A_{\gamma p}$ and $B_{\gamma p}$ are constants.

The quality of the fits is so good, with chi-squared of less than one by d.o.f., that 
one may consider studying the corrections to (1.4), and the interpolation between (1.4)  
and (1.5); this we do in Sect.4. In Sect.5 we discuss the ultra high momentum limit of our 
formulas, in particular in connection with the BFKL approach which suggests the asymptotic 
behaviour
$$F_2(x,Q^2)\sim x^{c_0\alpha_s}.\eqno (1.6)$$
We study the region of validity of (1.4), limited 
for fixed $x$ by a certain $Q^2(x)$. We get indications that, for 
$x\lsim 10^{-2}$, $Q^2(x)\sim 200\,-\,300\;{\rm GeV}^2$.

 In what regards the 
transition from (1.4) to (1.6) we get (not conclusive, however) evidence 
against it at least in the region covered by the HERA data, i.e.,  
 below 
$Q^2\simeq 800\;{\rm GeV}^2$. 
\vfill\eject
\noindent 2.-{\bf DEEP INELASTIC SCATTERING AT }$x\rightarrow 0$
\vskip.3cm
2.1.-{\sl General considerations}
\vskip.5cm
For DIS the relevant quantities are the structure functions,
$$F_2(x,Q^2),\;F_G(x,Q^2);
\;R(x,Q^2)=\frac{F_2(x,Q^2)-F_1(x,Q^2)}{F_2(x,Q^2)}.\eqno (2.1)$$
$F_G$ is the gluon structure function, and $R$ is (proportional to) the 
longitudinal one.

It is convenient to use the singlet function $F_S$, 
normalized so that the momentum sum rule reads
$$\int^1_0\dd x [F_S(x,Q^2)+F_G(x,Q^2)]
\rightarrowsub_{Q^2\rightarrow\infty} 1.\eqno (2.2)$$
The relation between $F_2$ and $F_S$ is as follows: one has
$$F_2(x,Q^2)=\langle e^2_q\rangle \left[F_S(x,Q^2)+F_{NS}(x,Q^2)\right]\,,\eqno (2.3{\rm a})$$
where $F_{NS}$ is the  so-called 
nonsinglet structure function. Since in the limit $x\rightarrow 0$ 
it decreases very fast compared to $F_S$, we may consider the approximate relation 
$$F_2(x,Q^2)\simeqsub_{x\rightarrow 0}\langle e^2_q\rangle F_S(x,Q^2).\eqno (2.3{\rm b})$$
 The average charge of the excited flavours in 
(2.3) is $\langle e^2_q\rangle=\tfrac{5}{18}$ for $n_f=4$ or 
 $\langle e^2_q\rangle=\tfrac{11}{45}$ for $n_f=5$. For the values 
of $Q^2,\,x$ with $8\; {\rm GeV}^2\lsim Q^2\lsim 65 \;{\rm GeV}$ we 
are in a mixed situation in which bottom is excited in the $s$ variable but 
not in the $Q^2$ variable. We will thus present results for both $n_f=4\;{\rm and}\;5$: 
as will be seen they differ very little. For $Q^2\lsim 9 \;{\rm GeV}^2$ only 
three flavours are excited in the $Q^2$ variable. Both 
for this and other reasons this energy region requires a specific treatment. 

The evolution equations for the structure functions
 may be written in two ways. One has the so called Altarelli-Parisi, or DGLAP 
equations$^{[2,3]}$, that we write only for 
the singlet structure functions, 
$$\frac{\partial F_i(x,Q^2)}{\partial t}=\sum_j\int^1_x\dd z\,P_{ij}(z)F_j(x/z,Q^2),
\eqno (2.4)$$
$$i,j=S,G;t=\log Q^2,$$
and the explicit form of the splitting functions $P_{ij}$ may be found 
in refs. 1,3. Alternatively, one may define the {\it moments} of the 
structure functions,
$$\matrix{\mu_S(n,Q^2)=\int^1_0\dd x\,x^{n-2}F_S(x,Q^2),\cr
\mu_G(n,Q^2)=\int^1_0\dd x\,x^{n-2}F_G(x,Q^2)\cr}\eqno (2.5)$$
and then the integro-differential equations (2.4) may be solved for 
the $\mu_i$:
$$\left(\matrix{\mu_S(n,Q^2)\cr \mu_G(n,Q^2)\cr}\right)=
\left[\frac{\alpha_s(Q_0^2)}{\alpha_s(Q^2)}\right]^{{\bf D}(n)}
\left(\matrix{\mu_S(n,Q_0^2)\cr \mu_G(n,Q_0^2),\cr}\right) \eqno (2.6{\rm a})$$
where the {\it anomalous dimension matrix} ${\bf D}(n)$ is$^{[4]}$
\vfill\eject
$${\bf D}(n)=\frac{16}{33-2n_f}$$
$$\times\left(\matrix{\dfrac{1}{2n(n+1)}+\frac{3}{4}-S_1(n)
&\dfrac{3n_f}{8}\,\dfrac{n^2+n+2}{n(n+1)(n+2)}\cr
\dfrac{n^2+n+2}{2n(n^2-1)}&
\dfrac{9}{4n(n-1)}+\dfrac{9}{4(n+1)(n+2)}+\dfrac{33-2n_f}{16}-\dfrac{9S_1(n)}{4}\cr}
\right).\eqno (2.6{\rm b}) $$
The function $S_1$ is related to the digamma function,
$$S_1(z)=\psi(z+1)+\gammae,\;
\psi(z)=\dd \log \Gamma(z)/\dd z,\,\gammae=0.5772\dots.$$
It should perhaps be stressed that eqs. (2.4) and (2.6) are {\it strictly 
equivalent}, as known from the very first works on the subject$^{[3]}$: (2.6) 
follows from (2.4) by taking moments and integrating, and (2.4) from 
(2.6) by inverting the Mellin transform (2.5) and differentiating.

Eqs (2.6), in the limit $x\rightarrow 0$, were solved
 long ago by L\'opez and the present author for $F_{NS}$, 
 $F_2$ and $F_G$ 
 to leading order (LO) in ref.5; including next to leading corrections in 
ref.6 and, for the longitudinal structure function by the quoted authors 
and Gonz\'alez-Arroyo in ref.7 [where the interested reader 
may find the extension of eqs. (2.4), (2.6) to $R(x,Q^2)$]. The comparison 
with experiment was carried to leading order in refs. 5,7 ; the 
 comparison with experimental data
 of the calculation to next to leading order (NLO) was given in ref.8.

The LO results are remarkably simple. It was proved in ref.5 that, under 
certain conditions, the 
structure functions are given by the very explicit 
expressions, which therefore solve the evolution equations,
\footnote*{For ease of reference we will reproduce the proof in Sect.6 here.}
$$F_S(x,Q^2)\simeqsub_{x\rightarrow 0}
\hat{B}_S[\alpha_s(Q^2)]^{-d_+(1+\lambda)}x^{-\lambda},\eqno (2.8{\rm a})$$
$$F_G(x,Q^2)\simeqsub_{x\rightarrow 0}
\frac{B_G(1+\lambda)}{\langle e_q^2\rangle\hat{B}_S}\, F_S(x,Q^2),\eqno (2.8{\rm b})$$
$$F_{NS}(x,Q^2)
\simeqsub_{x\rightarrow 0}\hat{B}_{NS}[\alpha_s(Q^2)]^{-D_{11}(1-\rho)}x^{\rho}
,\eqno (2.8{\rm c})$$ 
$$R(x,Q^2)\simeqsub_{x\rightarrow 0}
r_0(1+\Lambdav)\frac{\alpha_s(Q^2)}{\pi}
.\eqno (2.8{\rm d})$$
We will also use the quantities
$$B_S\equiv\langle e_q^2\rangle\hat{B}_S,\;B_{NS}\equiv\langle e_q^2\rangle\hat{B}_{NS}.
\eqno (2.8{\rm e})$$
$\lambda$, $B_{NS}$ and $B_S$ are free parameters, independent of $x,Q^2$, although 
they may have a dependence on $n_f$ (expected to be slight; it will 
be discussed in  Sect.6). $\rho$ is known from Regge theory to be 
the intercept of the $f^0-\rho$ trajectory, $\rho\simeq0.5$. 
The quantities $B_G,r_0$ 
are obtainable in terms of $\lambda,\,B_S$:
$$\matrix{\displaystyle{\frac{B_G(1+\lambda)}{\hat{B}_S}=
\frac{d_+(1+\lambda)-D_{11}(1+\lambda)}{D_{12}(1+\lambda)};}\cr
{\displaystyle r_0(1+\lambda)=
\frac{4}{3(2+\lambda)}
\left[1+
\frac{3n_f}{2}\,\frac{d_+(1+\lambda)-D_{11}(1+\lambda)}{(3+\lambda)D_{12}(1+\lambda)}\right]}.
\cr}
\eqno (2.8{\rm f})$$
Here the $D_{11}(1-\rho)$, $D_{ij}(1+\lambda)$ and the $d_{\pm}(1+\lambda)$,  
$$d_{\pm}(1+\lambda)=\tfrac{1}{2}
\left\{D_{11}(1+\lambda)+D_{22}(1+\lambda)\pm 
\sqrt{[D_{11}(1+\lambda)-D_{22}(1+\lambda)]^2+
4D_{12}(1+\lambda)D_{21}(1+\lambda)} \right\}$$
are the matrix elements and eigenvalues of the matrix 
${\bf D}(n)$ evaluated (respectively) at $n=1-\rho$, $n=1+\lambda$ (the eigenvalues ordered so that 
$d_+>d_-$). We note that eqs. (2.5), (2.6) are valid for 
arbitrary (even complex) values of $n$. Thus, eqs. (2.8) give an explicit expression for 
all three structure functions $F_S,\,F_G$ and $R$ in terms of the only two free 
parameters $\lambda$ and $B_S$ ($B_{NS}$ also intervenes if
 we include $F_{NS}$ in the analysis).

Surprisingly enough, eqs. (2.8) and, more generally, the 
work of refs.5-8 seem to have been ignored by 
recent publications$^{[9,10]}$. Even more surprising, the HERA physicists have 
not used eqs. (2.8) in spite of their simplicity to analyse the experimental 
data, relying instead on a painstaking numerical integration of the
 Altarelli-Parisi equations (2.4) --a procedure not only infinitely 
more complicated than use of the explicit solutions (2.8), but that, 
as will be shown, produces larger errors.
\vskip.3cm
2.2.-{\sl Low $Q^2$ and medium $x$ analysis, and higher $Q^2$ and very small 
$x$ predictions}
\vskip.5cm
In the 1980 paper, L\'opez and the present author analyzed DIS data in the range
$$3.25\,{\rm GeV}^2\leq Q^2\leq 22.5\,{\rm GeV}^2,\;x>2\times10^{-2}.\eqno (2.9)$$
We found that the parameter $\lambda$ could be well determined while $B_S$ (which 
gives the overall normalization) was more uncertain. This uncertainty was due 
basically to the strong contribution of subleading effects, especially 
those due to the nonsinglet structure function, at the (relatively) large 
values of $x$ used. One had
$$\lambda=0.37\pm0.07,\;0.02\geq B_S\geq 0.001.\eqno (2.10{\rm a})$$
[Actually the value and error take into account also the analysis
 of $R$; $F_2$ alone would produce a slightly smaller $\lambda$,
 $\lambda=0.36\;$]. The results 
of the evaluation were subsequently shown to be essentially unaltered by 
inclusion of NLO corrections$^{[8]}$. As for $B_{NS}$, the fits based on 
functions containing both singlet and nonsinglet parts fix it only within large errors. 
Fortunately, however, the $W_3$ structure function in neutrino 
scattering is purely nonsinglet and allows a reasonably precise determination:  
$\hat{B}_{NS}\sim 1.4\;{\rm to}\;1.8$ or$^{[8]}$,
$$B_{NS}\simeq 0.3\;{\rm to}\;0.6.\eqno (2.10{\rm b})$$
With these numbers it is a trivial matter to {\it predict} the very low $x$ 
measurements of $F_2$ at HERA in terms of the only parameter $B_S$ that has 
to be taken as essentially free because, as 
explained before, it is not well fixed
 by the old, larger $x$ data. In a more precise determination 
we will of course also allow $\lambda,\,B_{NS}$ and even $\Lambdav$ to vary.

We will perform the analysis in two steps. In the first we will consider the old, 1993   
data$^{[11]}$, whose errors were typically $O(10\%)$, and with points 
in the range $x<10^{-2}, 8.5\,{\rm GeV}^2\leq Q^2\leq 65\,{\rm GeV}^2$; 
discussing then the more recent, as yet unpublished set of data which 
 have much smaller errors, and cover a wider range. Later in 
Sects.4, 5 we will consider the {\it theoretical} corrections to the formulas used. 

The results of the analysis of the 1993 data are presented
 in the following two tables. There  
 the value of $\lambda$ choosen is 
0.38, as it gives the best fit. The approximation of neglecting
 $F_{NS}$, i.e., Eq. (2.3b) was used. 
 For the QCD parameter $\Lambdav$ we first take it such that we reproduce the 
value of $\alpha_s(m_{\tau}^2)=0.32$, i.e., 
$$\Lambdav(n_f=4,1\,{\rm loop})=0.200\;{\rm GeV},\;
\Lambdav(n_f=5,1\,{\rm loop})=0.165\;{\rm GeV} .$$

\vskip.3cm
\hrule
\vskip.1cm
$$\matrix
{\lambda & d_+(1+\lambda) & B_S& B_G(1+\lambda)/B_S&r_0(1+\lambda)&\chi^2/{\rm d.o.f.}\cr
0.38\pm0.01&2.406\pm0.100&(2.70\pm0.22)\times 10^{-3}&20.56\pm0.54&6.24\pm 0.24&9.13/(32-2)\cr
}$$

\vskip.1cm
\centerline{{\bf Table I}{\it a}.- $n_f=4;\Lambdav(1\,{\rm loop},n_f=4)=0.200\,{\rm GeV};
\,\alpha_s(m_{\tau}^2)=0.32.$}
\vskip.2cm
\hrule
\vskip.1cm
\hrule
\vskip.3cm

$$\matrix
{\lambda & d_+(1+\lambda) &B_S& B_G(1+\lambda)/B_S&r_0(1+\lambda)&\chi^2/{\rm d.o.f.}\cr
0.38\pm0.01&2.595\pm0.110&(2.19\pm0.11)\times 10^{-3}&18.53\pm0.57&6.20\pm 0.29&9.34/(32-2)\cr
}$$

\vskip.1cm
\centerline{{\bf Table I}{\it b}.- $n_f=5;\Lambdav(1\,{\rm loop},n_f=5)=0.165\,{\rm GeV};
\,\alpha_s(m_{\tau}^2)=0.32.$}
\vskip.2cm
\hrule
\vskip.1cm
\hrule
\vskip.3cm

\setbox1=\vbox{\hsize80mm \epsfxsize=8.truecm\epsffile{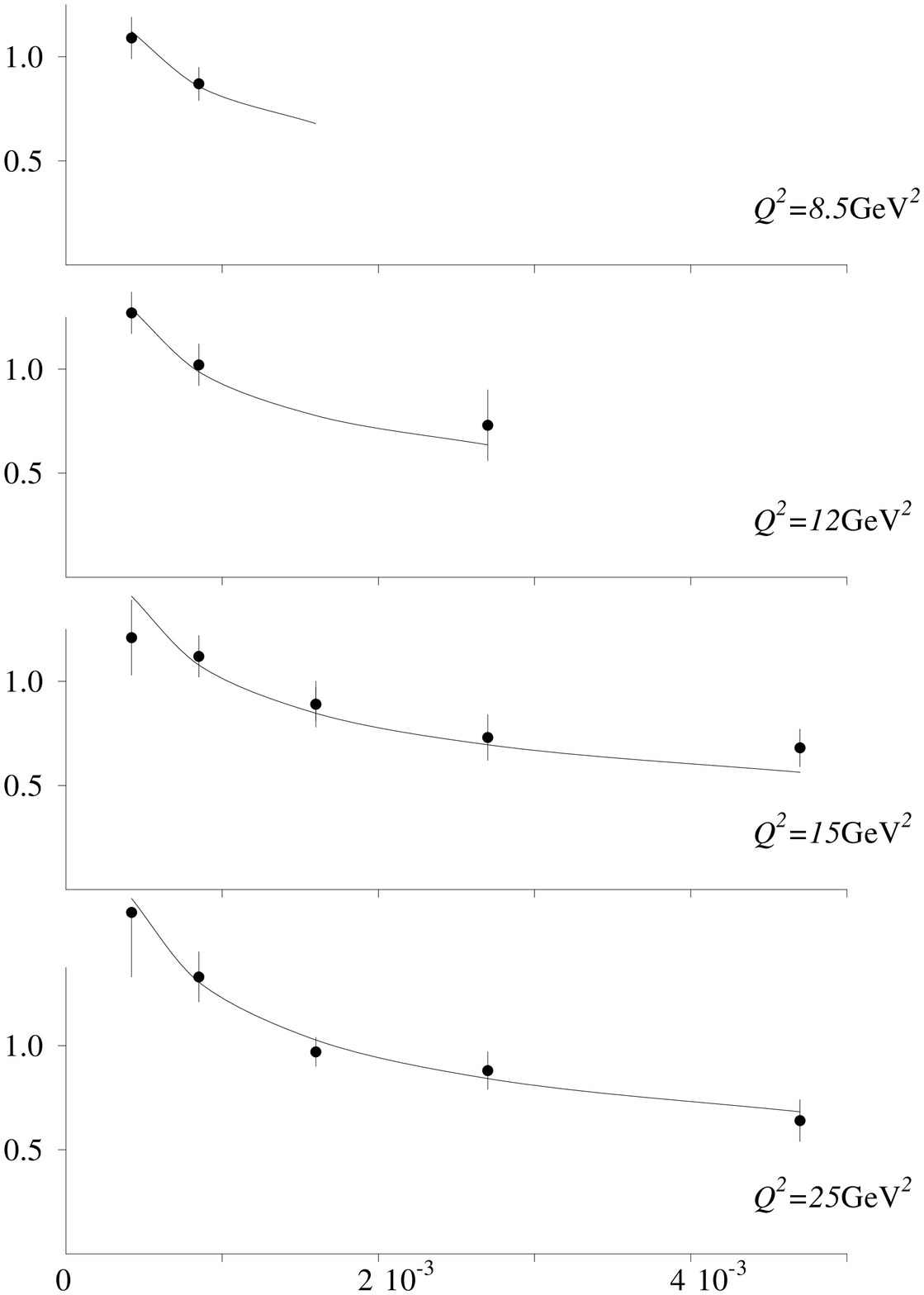}}
\setbox2=\vbox{\hsize80mm \epsfxsize=8.truecm\epsffile{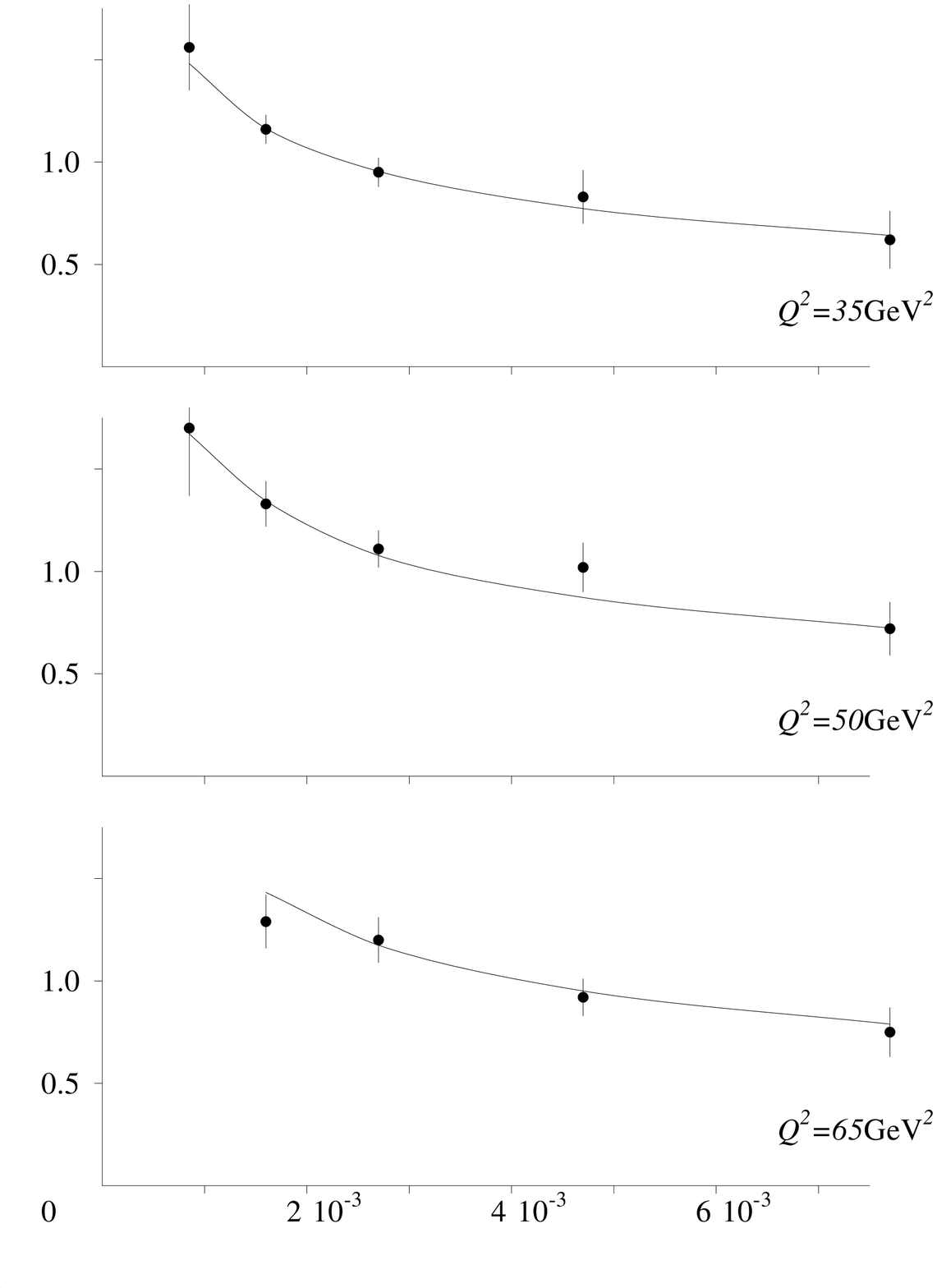}}
\line{\box1\hfill\box2}
\vskip.3cm
\centerline{{\bf Fig.}1.- Comparison of theoretical prediction, eq. (2.11) with 
experiment.}
\vskip.2cm
One can consider fitting also the QCD parameter, $\Lambdav$. In this case one 
discovers that, due to the slow, logarithmic variation of 
$\alpha_s$ with $\Lambdav$ and the large size of the experimental errors, the 
 effect of altering $\Lambdav$ may be largely compensated by a change in $B_S$. As 
an indication, we give the results of an evaluation with a small 
$\Lambdav$:

\vskip.3cm
\hrule
\vskip.3cm
$$\matrix
{\lambda & d_+(1+\lambda) &B_S& B_G(1+\lambda)/B_S&r_0(1+\lambda)&\chi^2/{\rm d.o.f.}\cr
0.35\pm0.01&2.735&1.19\times 10^{-3}&22.53&6.93&7.88/(32-2)\cr
}$$
\centerline{{\bf Table II}.- $n_f=4;\Lambdav(1\,{\rm loop},n_f=4)=0.10\,{\rm GeV};
\,\alpha_s(m_{\tau}^2)=0.28.$}
\vskip.2cm
\hrule
\vskip.1cm
\hrule

\vskip.3cm
The difference between these results and those of Tables I{\it a,b} are to be considered as an 
added uncertainty in all that follows. For the sake of conciseness, we 
will however consider most of the time only the parameters reported in 
Table I{\it a}. The graphic comparison with experiment is shown in Fig.1.
Here, 
because of the smallness of the $\chi^2$, we have only given the
 curve for the central values,
$$F_2(x,Q^2)=(2.70\times 10^{-3})\, [\alpha_s(Q^2)]^{-2.406}x^{-0.38},\;
\alpha_s(Q^2)=\frac{12\pi}{(33-2n_f)\log Q^2/(0.2\,{\rm GeV})^2}.\eqno (2.11{\rm a})$$
Morover, the error bars represent the statistical plus 
systematic errors composed quadratically.

We find the phenomenon already encountered in the analysis of the old, low energy data:
 $\lambda$ is well determined, but $B_S$ is
 less precisely fixed. Nevertheless, the errors have 
diminished substantially, and the consistency between the present
 analysis and that of refs. 5--8 is, to say the least, remarkable.

\vskip9.5cm
\setbox1=\vbox{\hsize80mm \epsfxsize=80truemm\epsffile{ 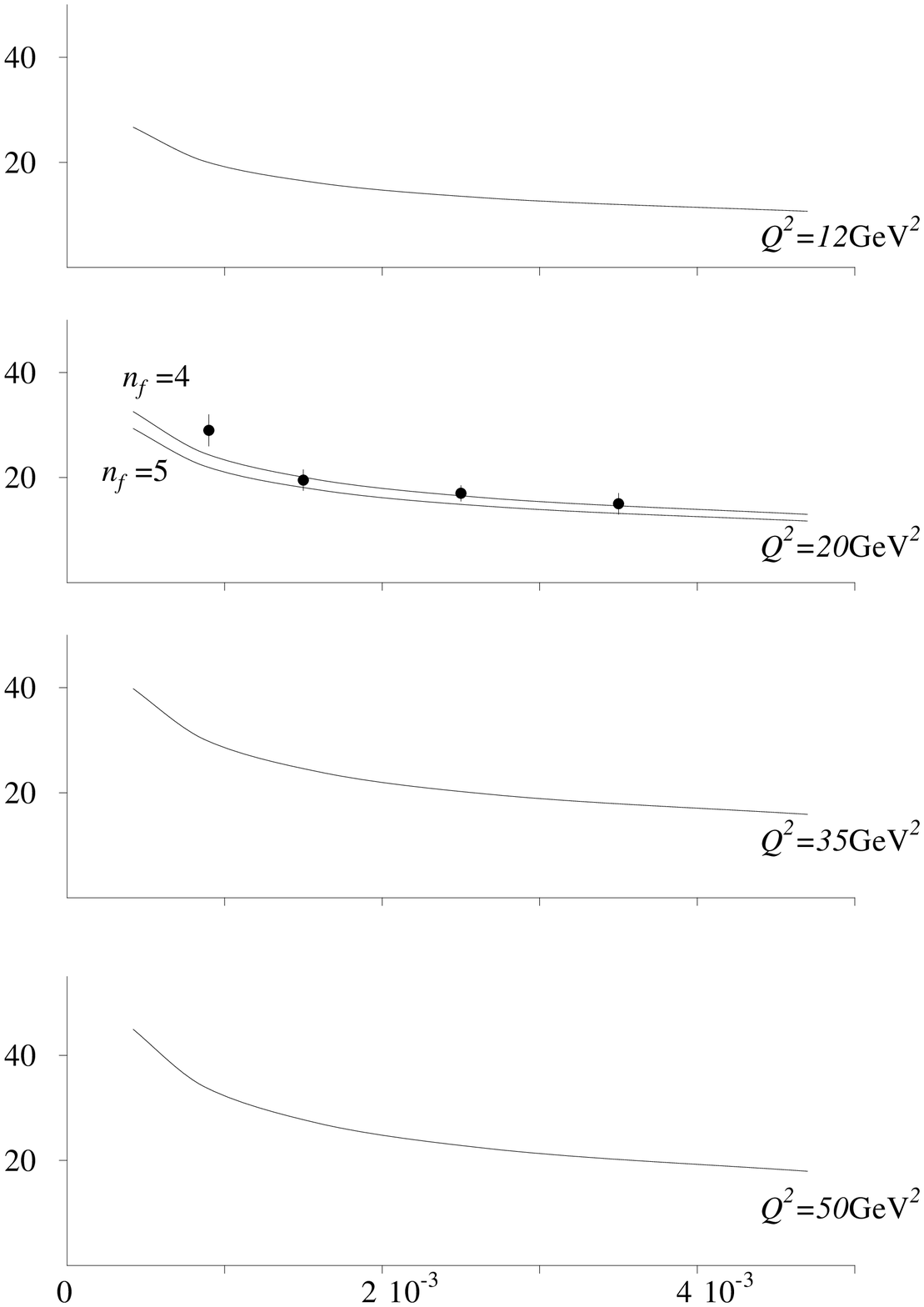}}
\setbox2=\vbox{\hsize 63mm{\noindent{\bf Figure} 2.- Prediction for the gluon structure 
function with $n_f=4$. For $Q^2=20$ GeV$^2$ also the 
evaluation with $n_f=5$ is shown. The points at that value of $Q^2$ are 
the results of the DGLAP calculation, {\it with statistical errors only.}\hb\vskip.3cm }}
\line{\box1\hfill\box2}
\vskip.3cm
\noindent 
The results obtained from the fit to $F_2$ 
permit us to calculate also, and without any extra parameter,
 the two remaining structure functions, which for the central values of the 
parameters are,
$$F_G(x,Q^2)=(55.5\times 10^{-3})\, [\alpha_s(Q^2)]^{-2.406}x^{-0.38},\eqno (2.11{\rm b})$$
$$R=6.24\frac{\alpha_s(Q^2)}{\pi}.\eqno (2.11{\rm c})$$
with $\alpha_s(Q^2)$ still as in (2.11a).

We do not plot $R$ as there are
 unfortunately no experimental data available. The gluon structure 
function is depicted in Fig.2, where for the sake of 
comparison we also show the results based on a numerical evaluation$^{[11]}$ using 
experimental data for $F_2$ and the 
DGLAP equations with the method of ref.10. The superior accuracy and simplicity of 
our formulas should be evident: note that the largest source of error 
in our evaluation is the uncertainty in the number of  flavours between 4 and 5.
\vfill\eject
2.3.-{\sl The new HERA results.}
\vskip.5cm
We will here consider the comparison of our predictions with 
the more recent HERA data (as yet unpublished). If one replaces, {\it tels quels}, 
the formulas used for the fits to the old data, then a very large chi-squared is 
produced. The reasons for this, rather obvious, are two. First of 
all, the precision of the data is such that the approximation of 
neglecting the $F_{NS}$ piece is no more justified. Secondly, the {\it range} 
is now such that the one loop evolution of $\alpha_s$ is not sufficiently accurate. 
If we restrict our analysis to $x<10^{-2}$, we have 63 data points with $Q^2$ varying in 
the bounds  
$12\;{\rm GeV}^2\leq Q^2 \leq 350\;{\rm GeV}^2$. We do not consider in the same fit  
smaller values of $Q^2$ because this would imply a
 large variation of $n_f$, on which $\lambda,\,B_S$ are 
expected to depend, and because the NLO corrections would certainly play an important role. 

We consider two possibilities: a restricted fit, only in the region
$$12\;{\rm GeV}^2\leq Q^2 \leq 90\;{\rm GeV}^2,$$
where it is sufficient to include the $NS$ contribution, and an evaluation in the full 
range where the two loop expression for $\alpha_s$ has to be taken into account as 
well.
\vskip.3cm
\hrule
\vskip.1cm
$$\matrix
{\lambda& B_{NS} &B_S&r_0&\chi^2/{\rm d.o.f.}\cr
0.39\pm0.01&1.03&2.73\times 10^{-3}&6.03&44.2/(48-3)\cr}$$
\vskip.1cm
\centerline{{\bf Table III}.
- $n_f=4;\,x<10^{-2};\,12\;{\rm GeV}^2\leq Q^2 \leq 90\;{\rm GeV}^2$
;$\;\Lambdav(1\,{\rm loop},n_f=4)=0.2\,{\rm GeV}$}
\vskip.2cm
\hrule
\vskip.1cm
\hrule
\vskip.3cm 
The results of the fits are again excellent. For the first 
case they are summarized in Table III, where we also give 
the parameters pertinent to the NS contribution, and 
$\Lambdav$ is considered {\it fixed}. For the full range, 
the results are given in Table IV. Note that 
the value we obtained of $\Lambdav$, taken now as a free parameter
 in the fit, {\it cannot} be considered 
as a true determination, as we have not included second order
 effects (that we discuss below).
\vskip.3cm
\hrule
\vskip.1cm
$$\matrix
{\lambda&\Lambdav(2\,{\rm loops},n_f=4)&d_+(1+\lambda) & B_{NS} &B_S&r_0&\chi^2/{\rm d.o.f.}\cr
0.38\pm0.01&0.10\pm0.01\,{\rm GeV}&2.406&0.772&(1.0\pm0.22)\times 10^{-3}&6.24&50.41/(63-4)\cr}$$
\vskip.1cm
\centerline{{\bf Table IV}.
- $n_f=4;\,x<10^{-2};\,12\;{\rm GeV}^2\leq Q^2 \leq 350\;{\rm GeV}^2$}
\vskip.2cm
\hrule
\vskip.1cm
\hrule
\vskip.3cm
\noindent We do not give the ensuing expressions for $F_G$ or $R$. If we take the values 
of the parameters, and the range of the variables pertaining to Table III, they are like 
 those in eqs. (2.11), {\it mutatis mutandi}. If we consider the 
situation in which we take $\alpha_s$ to two loops, Table IV, then the expressions for 
$F_G$, $R$ would not be more reliable than before because
 the two loop corrections to these quantities 
differ from the ones to $F_S$. Thus we give only the expression for the structure 
function $F_2$. We 
have,
$$F_2(x,Q^2)=(1.0\pm0.22)\times 10^{-3} [\alpha_s(Q^2)]^{-2.406}x^{-0.38}
+0.77[\alpha_s(Q^2)]^{-0.514}x^{0.5},\eqno (2.12{\rm a})$$
where now
$$\alpha_s(Q^2)=\frac{12\pi}{(33-2n_f)}\left\{1-6\frac{153-19n_f}{(33-2n_f)^2}
\,\frac{\log\log Q^2/\Lambdav^2}{\log Q^2/\Lambdav^2}\right\},\;n_f=4,\,\Lambdav=0.10\pm0.01.
\eqno (2.12{\rm b})$$
The comparison with experiment of the results from the restricted fit 
using the values of the parameters recorded in Table III woud give something very much 
like one what sees in the Figures 1 and 2. The comparison of the theoretical 
evaluation for $F_2$ using eq. (2.12) and experiment is shown in Fig.3.

\setbox1=\vbox{\hsize88mm \epsfxsize=88truemm\epsffile{ 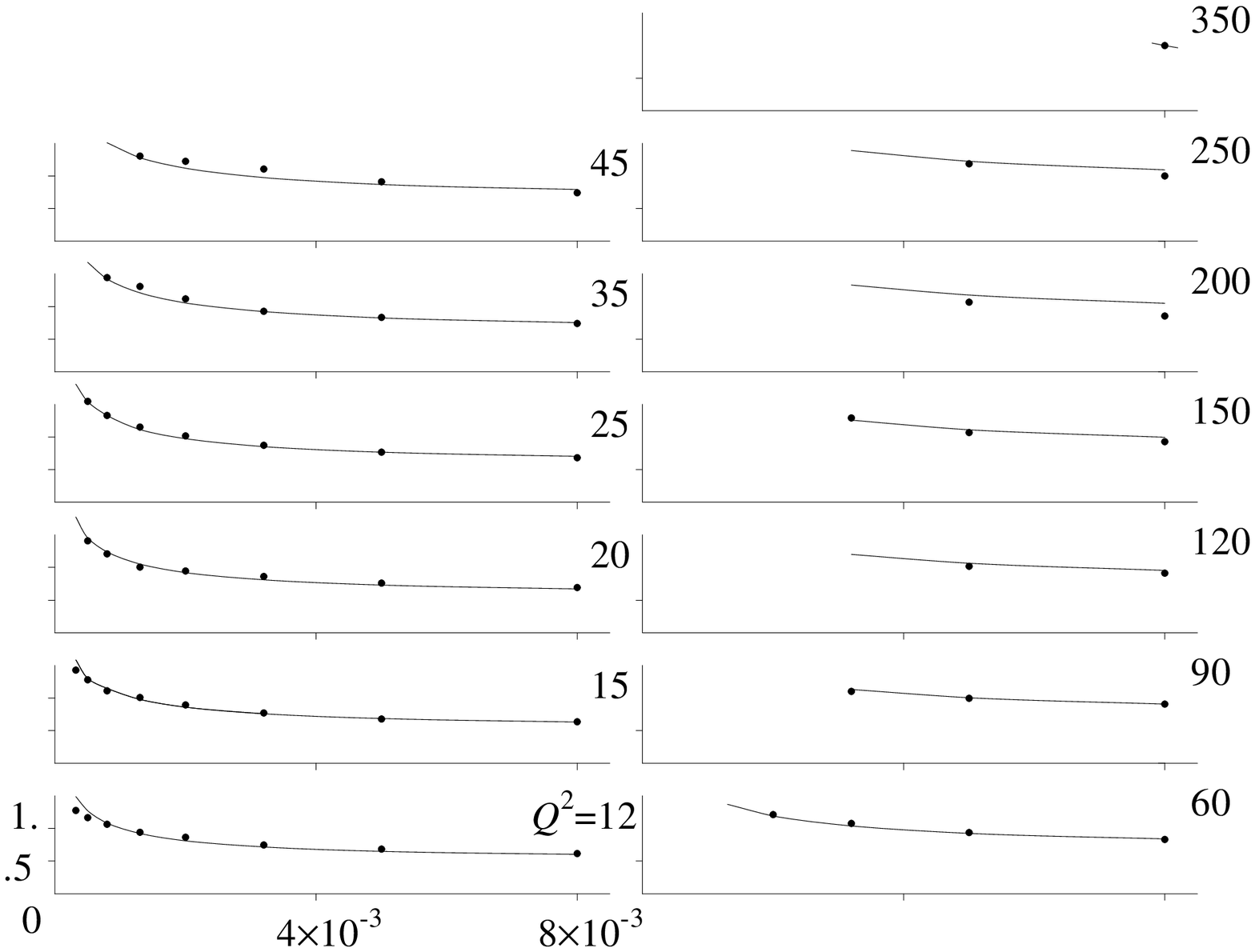}}
\setbox2=\vbox{\hsize 56mm{{\bf Figure} 3.- Fit to the latest 
\hb
\phantom{x}\kern1.5em HERA data with eq. (2.12) \hb
\phantom{x}\kern1.5em and $Q^2$ from 12 to 350  \gev$^2$. \hb
\phantom{x}\kern1.5em The errors of the experimental\hb
\phantom{x}\kern1.5em points, of the order of  the size \hb
\phantom{x}\kern1.5em of the dots, are not shown.\hb
\vskip.8cm }}
\line{\box1\hfill\box2}
\vskip.3cm 
 
To end this section we fit separately the low $Q^2$ data
 points, with a two loop expression for $\alpha_s$. We then find the results of 
Table V.
\vskip.3cm
\hrule
\vskip.1cm
$$\matrix
{\lambda&\Lambdav(2\,{\rm loops},n_f=3) & B_{NS} &
 B_S&\chi^2/{\rm d.o.f.}\cr
0.29\pm0.02&0.14\,{\rm GeV}&0.75&0.3\times 10^{-3}
&18/(45-3)\cr}$$

\vskip.1cm
\centerline{{\bf Table V}.
- $n_f=3;\,x<10^{-2};\,1.5\;{\rm GeV}^2\leq Q^2 \leq 8.5\;{\rm GeV}^2$}
\vskip.2cm
\hrule
\vskip.1cm
\hrule
\vskip.3cm
We do only give errors on $\lambda$. In fact the results in Table V should 
 be taken as representing {\it effective} estimates. 
The reason is that the values of $Q^2$ 
are so low that an exact treatment of subleading corrections would be 
essential to get realistic values of the parameters. This includes not only 
$O(\alpha_s)$ corrections, but diffractive ones as well. This 
may easily alter substantially the results reported here; one should realize, for 
example, that even without the 
corrections a chi-squared/d.o.f. of less than one may be obtained provided 
$\lambda\lsim 0.35$, with $B_S\sim0.6\times 10^{-3}$ and $B_{NS}\sim 1.1$.
Note also that the compatibility of the results reported in Tables IV and V is 
made more apparent if we compare values of the
 $\hat{B}_S,\,\hat{B}_{NS}$, i.e., we extract 
the  factor $\langle e^2_q\rangle=5/18$ $(n_f=4)$ or 2/9 ( for $n_f=3$). Then we get,
$$\hat{B}_S=3.6\times 10^{-3}\; (Q^2>10\,{\rm GeV}^2);
\;(1.4\;{\rm to}\;2.7)\times 10^{-3} (Q^2<10\,{\rm GeV}^2);$$
$$\hat{B}_{NS}=3.11\; (Q^2>10\,{\rm GeV}^2);\;3.4\;{\rm to}\;4.4\; (Q^2<10\,{\rm GeV}^2).$$

\vfill\eject
\noindent 3.-{\bf HIGH ENERGY COMPTON SCATTERING}
\vskip.3cm
The results of the previous section indicate that a behaviour
$$F_2(x,Q^2)\simeqsub_{x\rightarrow 0}f_S(Q^2)s^{\lambda}+f_{NS}(Q^2)s^{0.5}+\dots
\eqno (3.1)$$
where $s$ is the c.m. energy squared, and we set the scale so that it is measured in 
GeV$^2$, may hold beyond the region of applicability of
 perturbative QCD; note that $s\simeq Q^2/x$. This was first suggested in 
ref.12 for Compton scattering, $\gamma+p\rightarrow {\rm all}$. The supposed subleading terms, 
denoted by dots in (3.1) were identified with diffractive effects --the Pomeron. 
To be precise, one may suppose that the photon has a ``hard", or pointlike component to 
which (3.1) applies (without the dots); and a 
``soft", hadronic or Pomeron component, which in old fashioned 
photon physics$^{[13]}$ was identified as connected 
with the probability of finding the rho 
resonance in the photon. One thus writes
$$\sigma_{\gamma p}(s)=\sigma^h_{\gamma p}(s)+\sigma^P_{\gamma p}(s), \eqno (3.2{\rm a})$$
where
$$\sigma^h_{\gamma p}(s)=B_{\gamma p}(s/1\,{\rm GeV}^2)^{\lambda},\eqno (3.2{\rm b})$$
$$\sigma^P_{\gamma p}(s)=A_{\gamma p}\hat{\sigma}^P(s)\eqno (3.2{\rm c})$$
and $\hat{\sigma}^P_{\gamma p}(s)$ is a universal, Pomeron hadronic 
cross section [into which, for reasons of 
convenience, we have also added the $\rho-f^0$ trajectory contribution, 
 corresponding to the term 
$f_{NS}(Q^2)s^{0.5}$ in (3.1)]. 

In view of its universality, $\hat{\sigma}^P(s)$ may be obtained, up to 
a constant, from any hadronic scattering cross sections, say
$$\sigma_{\pi p}(s)=C_{\pi p}\hat{\sigma}^P(s);\;
\sigma_{\bar{p}-p; p}(s)=C_{\bar{p}-p; p}\hat{\sigma}^P(s).\eqno (3.3)$$
The formulas employed for fitting the various {\it hadronic} cross sections are thus, 
$$\sigma_{hp}(s)=C_{hp}\hat{\sigma}(s),\eqno (3.3)$$
where the constant $C_{hp}$ depends on the process and $\hat{\sigma}(s)$ 
is obtained saturating the Froissart bound in the improved version of ref.14\footnote*{
This corrects an error in ref.14, where the scale factor is wrongly given as 
$\log^7(s/m_{\pi})$ instead of the correct value
 $\log ^{{\scriptscriptstyle 7}\over{\scriptscriptstyle2}} (s/m_{\pi}^2)$, cf. (3.4) below.} 
(but with the overall constant and an additive constant left as 
free parameters), plus a Regge pole contribution corresponding to the 
$f^0$ trajectory, which as explained 
we have found it convenient to incorporate into $\hat{\sigma}^P$:
$$\hat{\sigma}(s)=
A_F\log\frac{s}{m_{\pi}^2\log ^{{\scriptscriptstyle 7}\over{\scriptscriptstyle2}} (s/m_{\pi}^2)}
+1+A_{f^0}(s/1{\rm GeV}^2)^{-0.5}.\eqno (3.4)$$
The values of $A_F,\,A_{f^0}$ and $C_{hp}$ are obtained fitting the cross sections 
$\sigma_{\pi p}\equiv\frac{1}{2}[\sigma_{\pi^+ p}+\sigma_{\pi^- p}]$ and 
$\sigma_{Np}\equiv\frac{1}{2}[\sigma_{\bar{p} p}+\sigma_{pp}]$. For 
energies above 500 GeV we assume $\sigma_{\bar{p} p}=\sigma_{pp}+{\rm ``Regge"}$, 
where the piece ``Regge" is obtained extrapolating the low energy difference
 $\sigma_{\bar{p} p}-\sigma_{pp}$ to high energy with a Regge pole formula,
 $\sigma_{\bar{p} p}-\sigma_{pp}\simeq Cs^{\alpha_{\rho}}$.
 This gives a minute correction, 0.6 to 0.2 $mb$,
 for $s^{{\scriptscriptstyle 1}\over{\scriptscriptstyle 2}}
\simeq 500\;{\rm to}\;2000$ GeV.
 One then finds the universal parameters
$$A_F=0.0116,\,A_{f^0}=0.69\eqno (3.5{\rm a})$$
and moreover
$$C_{\pi p}=23.0\,{\rm mb};\;C_{Np}=39.6\,{\rm mb}.\eqno (3.5{\rm b})$$
Note that the ratio $C_{Np}/C_{\pi p}=1.72$ is reasonably close to the value 3/2  
predicted by the naive quark model.

It should be stressed that the present paper is {\it not} 
about hadronic cross sections. If a set 
of experimental measurements of $\pi p,\bar{p}p, pp$ existed from ``low" energies 
($s^{\frac{1}{2}}\sim 10\;{\rm GeV}$) to the large values reached in 
$\gamma p$ scattering, $s^{\frac{1}{2}}\sim 200\;{\rm GeV}$, a fit like that 
in eqs.(3.4,5) would be unnecessary; and indeed, one could also have used the 
phenomenological interpolations provided by the Particle Data Group$^{[15]}$. 
We choose (3.4) because of a theoretical prejudice in its
 favour, which is substantiated 
by the quality of the fits, with only two parameters, shown in Fig.4.

\setbox1=\vbox{\hsize80mm \epsfxsize=80truemm\epsffile{ 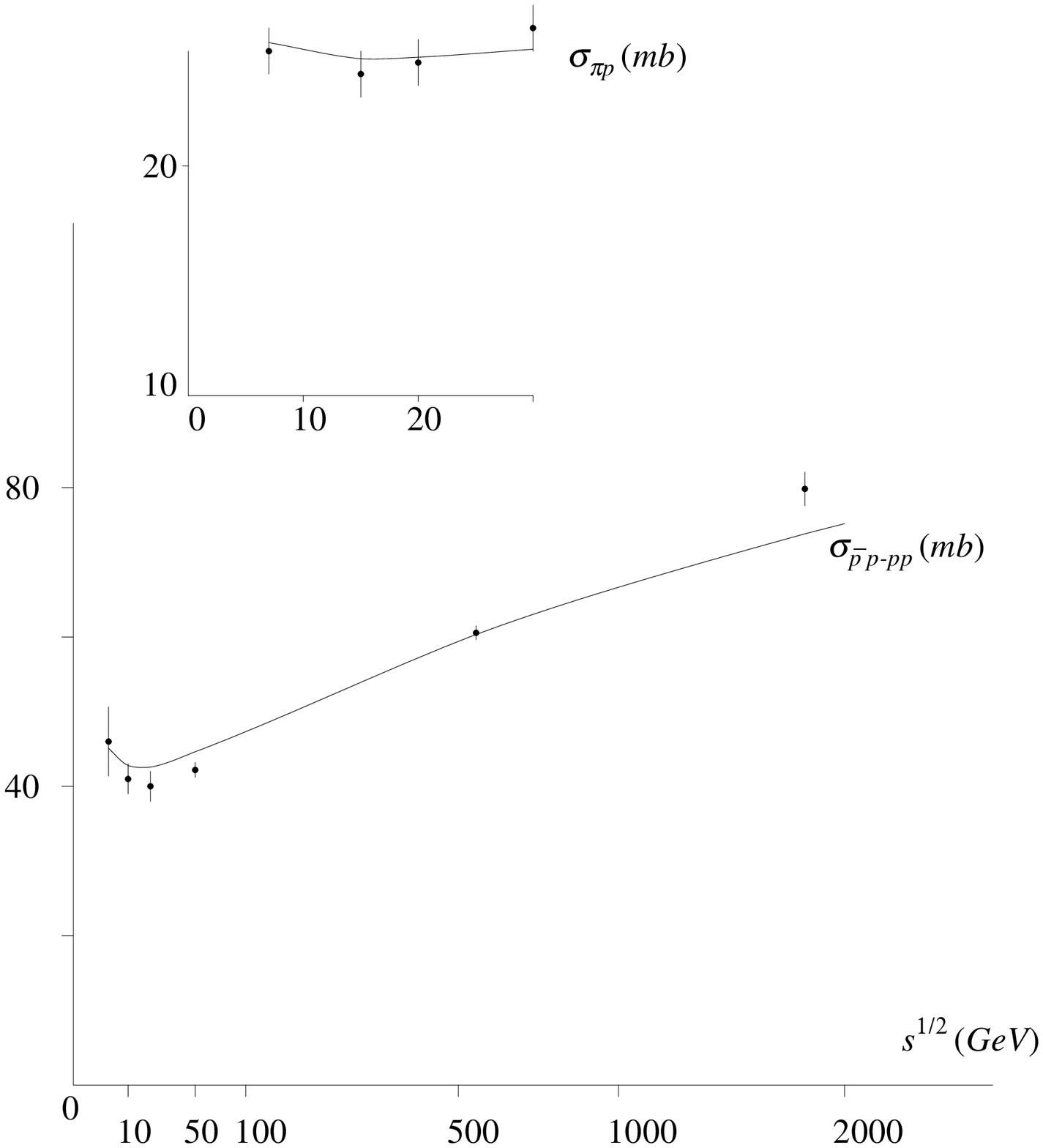}}
\setbox2=\vbox{\hsize 60mm{{\bf Figure} 4.- Fit to the 
$\pi p$ (indented)\hb
\phantom{x}\kern1.5em and $\tfrac{1}{2}[\bar{p}p+pp]$ cross 
sections with \hb
\phantom{x}\kern1.5em eqs. (3.4) and (3.5).\hb\vskip.3cm }}
\line{\box1\hfill\box2}
\vskip.3cm
With this we write $\sigma_{\gamma p}$ as in eqs. (3.2).  
The values of the constants one obtains are, for a $\chi^2/{\rm d.o.f.}=11.3/12$ (with 
statistical errors only),
$$C_{\gamma p}=102.0\pm 2.0\,\mu{\rm b};
\;B_{\gamma p}=1.64\pm0.30\,\mu{\rm b};\;\lambda=0.23\pm 0.03.\eqno (3.6)$$
The ratio $C_{\pi p}/C_{\gamma p}=225$ is close to the value 
220 obtained at low energy from experiment and vector meson dominance arguments$^{[12]}$.
 Finally, the parameters 
$\lambda=0.23$ and $B_{\gamma p}=1.64$ are reasonably similar to those obtained at 
``low energy" ($s^{\frac{1}{2}}\lsim 20\, GeV$) in ref.12, that is,  
 $\lambda=0.40$ and $B_{\gamma p}=1.40$, while $\lambda$ is also
 close to the value deduced in the previous subsection (2.3) from small $x$ in deep 
inelastic scattering, $\lambda\simeq 0.3$. One should not take the small discrepancies 
very seriously; the values recorded above for the parameters were obtained taking only 
{\it statistical} errors into account in the recent, very 
high energy HERA data$^{[16]}$. If we include also {\it systematic} errors, then 
we get $\lambda=0.26$ and a chi-squared/d.o.f. less than one provided 
$0.22\leq\lambda\leq 0.29$, perfectly compatible with what we found before, modulo 
the (expected; see  Sect.6) dependence of $\lambda$ on the 
number of excited flavours.

\setbox1=\vbox{\hsize80mm \epsfxsize=8truecm\epsffile{ 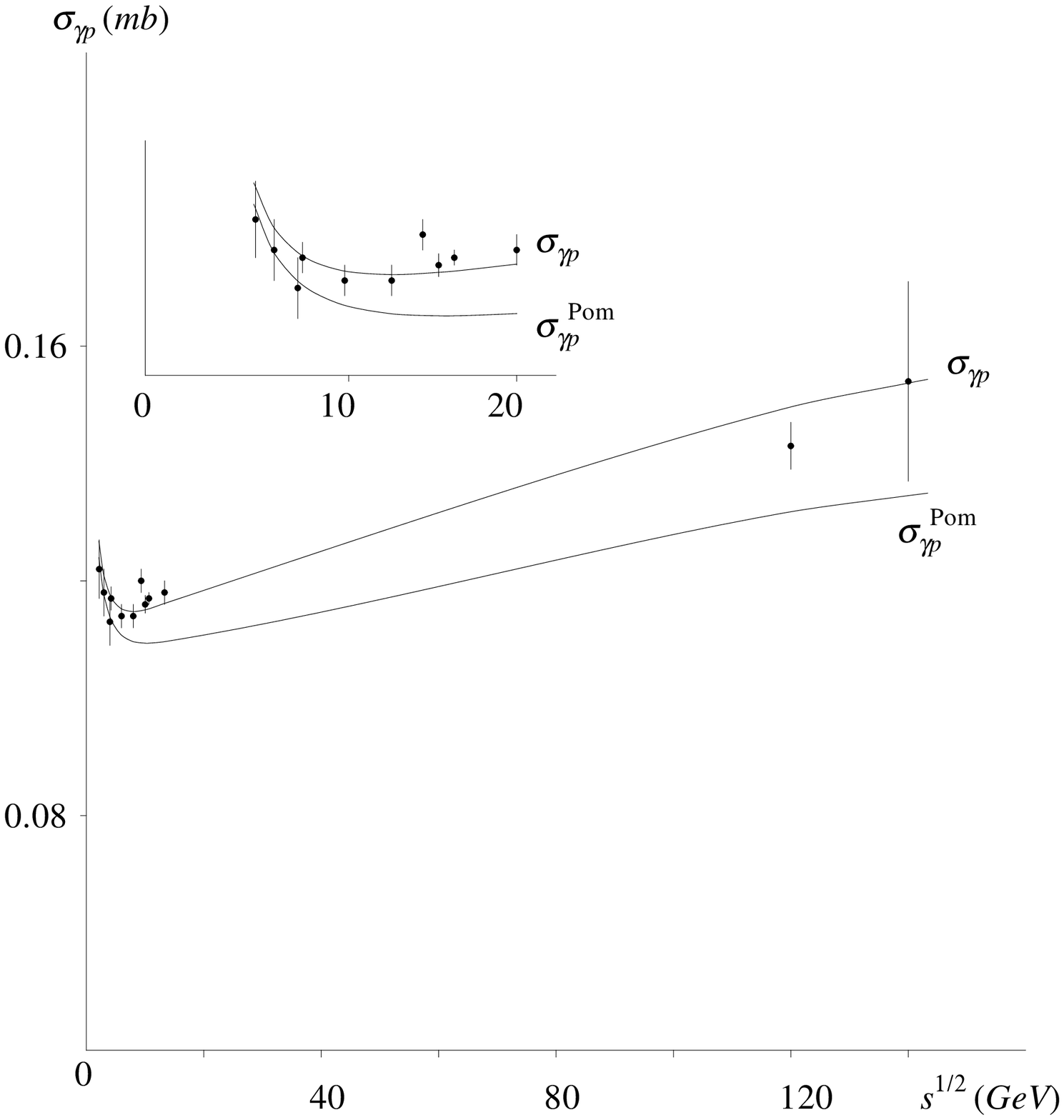}}
\setbox2=\vbox{\hsize 60mm{{\bf Figure} 5.- Fit to the 
$\gamma p$ \hb
\phantom{x}\kern1.5em  
 cross section with eqs.(3.2) \hb
\phantom{x}\kern1.5em In the indent, a blowup of \hb
\phantom{x}\kern1.5em the low energy region.\hb
\vskip1.7cm }}
\line{\box1\hfill\box2}
\vskip.3cm
The quality of the fit should be apparent from Fig. 5, where only statistical errors 
are shown for the high energy HERA data; also impressive is the 
stability of the extrapolation by one order of magnitude in energy from the 1980 
analysis to the present one.

\vfill\eject
\noindent 4.-{\bf SUBLEADING CORRECTIONS}
\vskip.3cm
4.1.-{\sl Two loop QCD corrections}
\vskip.5cm
The equations we have used take only into account LO (one loop) QCD, 
 except for the running of $\alpha_s(Q^2)$. A proper analysis should fully 
take into account NLO (two loop) corrections. These are known since the work of 
ref.6, and amount to replacing (2.8{\rm a-d}) by
$$F_S(x,Q^2)\simeqsub_{x\rightarrow 0}
\hat{B}_S\left[1+\frac{v_{S}(1+\lambda)\alpha_s(Q^2)}{4\pi}\right]
[\alpha_s(Q^2)]^{-d_+(1+\lambda)}x^{-\lambda},\eqno (4.1{\rm a})$$
$$F_G(x,Q^2)\simeqsub_{x\rightarrow 0}
B_G(1+\lambda)\left[1+\frac{v_{G}(1+\lambda)\alpha_s(Q^2)}{4\pi}\right] F_S(x,Q^2),\eqno (4.1{\rm b})$$
$$F_{NS}(x,Q^2)
\simeqsub_{x\rightarrow 0}\hat{B}_{NS}
\left[1+\frac{v_{NS}(1-\rho)\alpha_s(Q^2)}{4\pi}\right]
[\alpha_s(Q^2)]^{-D_{11}(1-\rho)}x^{\rho}
,\eqno (4.1{\rm c})$$
where the $v_i(n)$ are known, complicated combinations of one- and 
the two-loop anomalous dimension 
matrix elements and the one loop Wilson coefficients
 whose explicit expression may be found in ref.8. Because of this structure, the 
bulk of the NLO effect may be approximated {\it for a single 
structure function}, by just leaving $\Lambdav$ as a free parameter --precisely as we 
did in subsect. 2.3. The full NLO analysis, with eqs. (4.1) taken 
into account, requires more precision, in particular incorporating also other subleading 
effects and will be presented in a separate paper.
\vskip.3cm
4.2.-{\sl The joining of low and high $Q^2$}
\vskip.5cm
In Subsect. 2.3 and Sect. 3 we have shown that on can reproduce the 
cross sections
$$\sigma_{\gamma^*(-Q^2)p}(s)$$
for, respectively,  off shell (but with fairly small $Q^2$, down to 1.5 GeV$^2$)
 or on shell photons, $Q^2=0$, with formulas with  similar functional form 
for the structure function and the ``hard" piece of the Compton cross section. 
It is then tempting to try and join the two approaches. This presents a number of 
problems, and rises some very interesting questions, that we 
now briefly discuss, leaving the full 
analysis for a future publication.

First of all, it is clear that an equation like (2.8a) (say) cannot 
hold for $Q^2$ too small. Not only, if taken literally, it would 
become singular for $Q^2\sim\Lambdav^2$, but, from the relation between 
$F_2$ and $\sigma_{\gamma^*(-Q^2)p}(s)$ it follows that, as $Q^2\rightarrow 0$,
$$\sigma_{\gamma^*(Q^2) p}\simeqsub_{Q^2\rightarrow 0}
\frac{4\pi^2\alpha}{Q^2}\,F_2(x,Q^2),\eqno (4.2)$$
so $F_2(x,Q^2)$ must vanish proportional to $Q^2$ for $Q^2$ small.

There are three ways in which one may look at eqs.(2.8), (4.2). First of all 
comes the question of the joinig of the two. If we take literally the values 
of the parameters we found in Sect.2, and the {\it hard} part 
of the photon-proton cross section, this 
occurs for a very reasonable value of $Q^2$, $Q^2_{\rm crit}\sim 1\,{\rm GeV}$. 
It would not be difficult to write interpolation formulas, provided one 
tackles first the two remaining questions: the divergence of $\alpha_s(Q^2)$
for $Q^2\sim\Lambdav^2$, and what happens to the soft, or Pomeron 
contribution to $\sigma_{\gamma^*(-Q^2) p}$, numerically
 dominant when $Q^2=0$ and disappearing 
for $Q^2\gg\Lambdav^2$.
 
In what regards the matter of the divergence of $\alpha_s(Q^2)$ an interesting 
possibility is that, as has been suggested recently$^{[17]}$, $\alpha_s(Q^2)$ {\it freezes} 
at low $Q^2$. Specifically, if this idea is correct, one would have an expression 
for $\alpha_s$ like
$$\alpha_s(Q^2)=\frac{12\pi}{(33-2n_f)\log[(Q^2+M^2)/\Lambdav^2]}\eqno (4.3)$$
with $M^2$ a parameter that one expects to be of the order of 
a typical hadronic mass (some estimates give $M\sim 0.96\,{\rm GeV}$).
With respect to the remains of the Pomeron at large $Q^2$ one may 
return to its interpretation as the probability of finding a real rho
 inside a photon$^{[13]}$ and parametrize it with something like 
$$\sigma^P_{\gamma^*(-Q^2) p}\simeq
\left[\frac{m^2_{\rho}}{2m^2_{\rho}+Q^2}\right]^2\sigma^P_{\gamma (Q^2=0) p},$$
or perhaps adscribe the Pomeron to a diffractive effect and
 use instead an exponential form-factor:
$$\sigma^P_{\gamma^*(-Q^2) p}\simeq 
\ee^{-Q^2/b^2}\sigma^P_{\gamma (Q^2=0) p}.$$ 
  We will say no more about this, 
leaving the matter for the announced future publication.      
 \vfill\eject
\noindent 5.-{\bf VERY HIGH MOMENTA AND CONNECTION WITH THE BFKL CONJECTURE }
\vskip.3cm

The questions discussed in the previous section affect mostly the 
medium and low $Q^2$ behaviour of structure functions. We will
 now discuss a very important matter that becomes relevant at ultra high energies.
 
According to our equations we have (Table IV),
$$F_2(x,Q^2)\simeqsub_{x\simeq 0}B_S[\alpha_s(Q^2)]^{-d_+(1+\lambda)}\,x^{-\lambda}
,\eqno (5.1{\rm a})$$
$$\lambda= 0.38\pm0.01,\,d_+(1+\lambda)=2.406\;(n_f=4).\eqno (5.1{\rm b})$$
On the other hand, the so-called BFKL equations$^{[18]}$ imply
$$F_2(x,Q^2)\simeqsub_{x\simeq 0\atop Q^2\rightarrow \infty}
C_2\,x^{\omega_0\alpha_s(Q^2)},\;\omega_0=\frac{4C_A\log 2}{\pi},\;C_A=3.
\eqno (5.2)$$
Eqs. (5.1) and (5.2) are, on the face of it, incompatible. What 
is more, while (as we have seen) (5.1) agrees very well with experiment, 
(5.2) deviates from experimental data by dozens of standard deviations. 
 Since eq.(5.2) is only
 valid in the leading approximation and, at least to 
the author's knowledge, there is no control 
of the subleading corrections, one may be tempted to conclude that 
it should be rejected.

The situation, however, is not as clear as may appear at first sight. 
First of all, the behaviour of the anomalous dimension matrix 
${\bf D}(n)$ for large $n$ imply that, {\it for any 
fixed} $x$ the structure functions must tend to zero 
for very large $Q^2$, contrarily to the behaviour following 
from eqs.(5.1) which make them {\it grow} 
with  $Q^2$. Therefore, the behaviour (5.1) must stop at 
ultra large $x$--{\it dependent} values $Q^2\geq Q^2(x)$ where 
$F_2(x,Q^2)$ should start decreasing, and perhaps turn into something 
like (5.2). To be precise, the conditions 
of validity of (5.1) (for whose proof we refer to  Sect.6)
 are as follows. {\it Assume} that, at a given $Q^2_0$ sufficiently large 
for perturbation theory to apply, we have
$$F_2(x,Q_0^2)=f(Q_0^2)x^{-\lambda_0(Q_0^2)}+O(x^{-\mu}),\;\mu<\lambda_0(Q_0^2).$$
Then, for all $,Q^2\geq Q_0^2$, we have that, {\it for sufficiently small $x$}
$$F_2(x,Q^2)=({\rm Constant})[\alpha_s(Q^2)]^{-d_+(1+\lambda)}+O(x^{-\mu}),\;
\lambda=\lambda_0(Q_0^2)\;{\rm independent\;of\;}Q^2.$$
What the results of refs.5,6 do not say is starting from which 
$x=x(Q^2)$ is this behaviour valid. The analysis of experimental data carried 
so far indicates that, when $Q^2<350\;\gev^2$, the behaviour (5.1) 
holds for $x<10^{-2}$. In the remainder of this section we will 
address the question of what happens above such values of $x$ and $Q^2$.      

To clarify the situation, and make more quantitative the analysis, 
 we will consider the momentum sum rule [eq.(2.2)]. Separating 
the quark and gluon contributions one has (see e.g. ref.1),
$$\lim_{Q^2\rightarrow \infty}\int_0^1\dd x\,F_2(x,Q^2)
=\langle e^2_q\rangle\frac{3n_f}{16+3n_f}.\eqno (5.3)$$
If we saturate the sum rule with the asymptotic behaviour we 
find that (5.2) is compatible with (5.3) provided $C_2=
\langle e^2_q\rangle [3n_f/(16+3n_f)]$; but if 
we 
substitute (5.1), assuming it to be valid up to $x=x_0$, we get 
$$\int_0^{x_0}\dd x\,F_2(x,Q^2)\simeq\frac{B_S x_0^{1-\lambda}}{1-\lambda}[\alpha_s(Q^2)]^{-d_+(1+\lambda)}
\eqno (5.4)$$
which violates (5.3) for $Q^2>Q_{\rm lim}^2 (x_0)$ with $Q_{\rm lim}^2 (x_0)$ defined by
$$\alpha_s(Q_{\rm lim}^2 (x_0))=
\left\{\frac{B_S x_0^{1-\lambda}}{1-\lambda}\;\frac{16+3n_f}{3n_f\langle e^2_q\rangle}\right\}^{1/d_+}.$$     
  For $n_f=4$ and with the values of $B_S,\,\lambda$ we have found, say
$$B_S= 10^{-3},\,\lambda =0.38;\,\Lambdav=0.10$$
we get that sizeable ($\sim$ a few percent) corrections to (5.1) must occur when 
$x\sim 10^{-2}$ already for 
$$Q^2_{\rm lim}\sim 100\,{\rm GeV}^2.$$
 This means that the precision of the more recent HERA data may 
be able to reveal the corrections to (5.1), and even to discriminate 
against, or for, (5.2).

In order to be more quantitative about this, we start by assuming that at a fixed 
$Q_0^2$ one had, exactly,
$$F_S(x,Q_0^2)=\hat{B}_S[\alpha_s(Q_0^2)]^{-d_+(1+\lambda)}\,x^{-\lambda}
.\eqno (5.5)$$
Then the moments at $Q_0^2$ are
$$\mu_S(n,Q_0^2)=\frac{\hat{B}_S}{n-1-\lambda}[\alpha_s(Q_0^2)]^{-d_+(1+\lambda)}
,$$
and, in view of the evolution equations (2.6) we get the result, valid at 
all $Q^2$,
$$\mu_S(n,Q^2)=\frac{\hat{B}_S}{n-1-\lambda}[\alpha_s(Q^2_0)]^{-d_+(1+\lambda)}
\left[\frac{\alpha_s(Q_0^2)}{\alpha_s(Q^2)}\right]^{d_+(n)},\eqno (5.6)$$
an evaluation that holds to corrections of relative
 order $\alpha_s^{d_+-d_-}$. In particular, (5.6) is certainly compatible 
with the momentum sum rule in that, for $n=2$, $d_+(2)=0$ and thus 
$$\mu_S(2,Q^2)=\int^1_0\dd x\,F_S\rightarrowsub_{Q^2\rightarrow \infty}{\rm constant}.$$

Next we get the behaviour implied by (5.6) for $x\rightarrow 0$. The leading 
behaviour is still given by eqs. (2.8) [or (5.1)]. Because, however now 
(5.6) is assumed to be {\it exact}, we may
 find the corrections. These are dominated 
by the first singularity of $d_+(n)$, which occurs at $n=1$. After a simple calculation 
we find,
$$F_S(x,Q^2)=\hat{B_S}[\alpha_s(Q^2)]^{-d_+(1+\lambda)}x^{-\lambda}-\Deltav (x,Q^2)+O(x)
,\eqno (5.7{\rm a})$$
and $\Deltav (x,Q^2)$ is such that, for $n=1+\epsilon,\,\epsilon\rightarrow 0$, it 
is given by the leading singularity of $d_+(1+\epsilon)\simeq [36/(33-2n_f)\epsilon]$:
$$\frac{\hat{B}_S}{\lambda}[\alpha_s(Q_0^2)]^{-d_+(1+\lambda)}
\left[\frac{\alpha_s(Q_0^2)}{\alpha_s(Q^2)}\right]^{[36/(33-2n_f)\epsilon]}=
\int_0\dd x\,x^{\epsilon-1}\Deltav(x,Q^2).$$
At large $Q^2$ and small $x$ we have the asymptotic expression for the correction,
$$\Deltav (x,Q^2)\simeqsub_{x\rightarrow 0\atop Q^2\rightarrow\infty}
\frac{B_S}{\lambda\alpha_s(Q_0^2)^{d_+(1+\lambda)}}
\left[ \frac{9\log[(\log Q^2/\Lambdav^2)/(\log Q_0^2/\Lambdav^2)]}{(33-2n_f)\pi |\log x|}\,
\right]^{\frac{1}{2}}
\exp \sqrt{\frac{144|\log x|}{(33-2n_f)}\;\left[\log\frac{\log Q^2}{\log Q_0^2}\right]}.
\eqno (5.7{\rm b}) $$
We see how the momentum sum rule and the behaviour as $x\rightarrow 0$
 get reconciled. For $x\rightarrow 0$, the piece  
$\hat{B}_S[\alpha_s(Q^2)]^{-d_+(1+\lambda)}x^{-\lambda}$ 
dominates over $\Deltav(x,Q^2)$; but, for {\it fixed} $x$, $\Deltav$
 dominates when $Q^2\rightarrow\infty$ and in fact (5.7) cease to be valid. It 
is to be noted, however, that this mechanism {\it excludes} the BFKL conjecture (5.2) 
which is never attained.

We may generalize this last result, as the analysis depends only on the first 
singularity of ${\bf D}(n)$, located at $n=1$, occurring not after the 
exponent of the first correction to (5.5). So we get essentially the 
same result if, at a given $Q^2$ one only assumes
$$F_S(x,Q_0^2)=
\hat{B}_S[\alpha_s(Q_0^2)]^{-d_+(1+\lambda)}\,x^{-\lambda}+O(x^0).\eqno (5.8)$$
The conditions that one has the behaviour (5.2) then are that the correction to 
(5.1) be of the form $({\rm const.})x^{-\mu},\,\mu>0.$ In fact, it is not 
difficult to get convinced that would need an {\it infinite}
 set of terms so that one had,
$$\matrix{F_S(x,Q^2)=
\sum_{j=0}^{\infty}\hat{B}_{Sj}[\alpha_s(Q^2)]^{-d_+(1+\lambda_j)}x^{-\lambda_j}
-\Deltav(x,Q^2),\cr
\lambda_0=\lambda>\lambda_1>\lambda_2>\dots>0,\cr}
\eqno (5.9{\rm a})$$
$$\Deltav=\sum \Deltav_j. \eqno (5.9{\rm b})$$

\setbox1=\vbox{\hsize75mm \epsfxsize=75truemm\epsffile{ 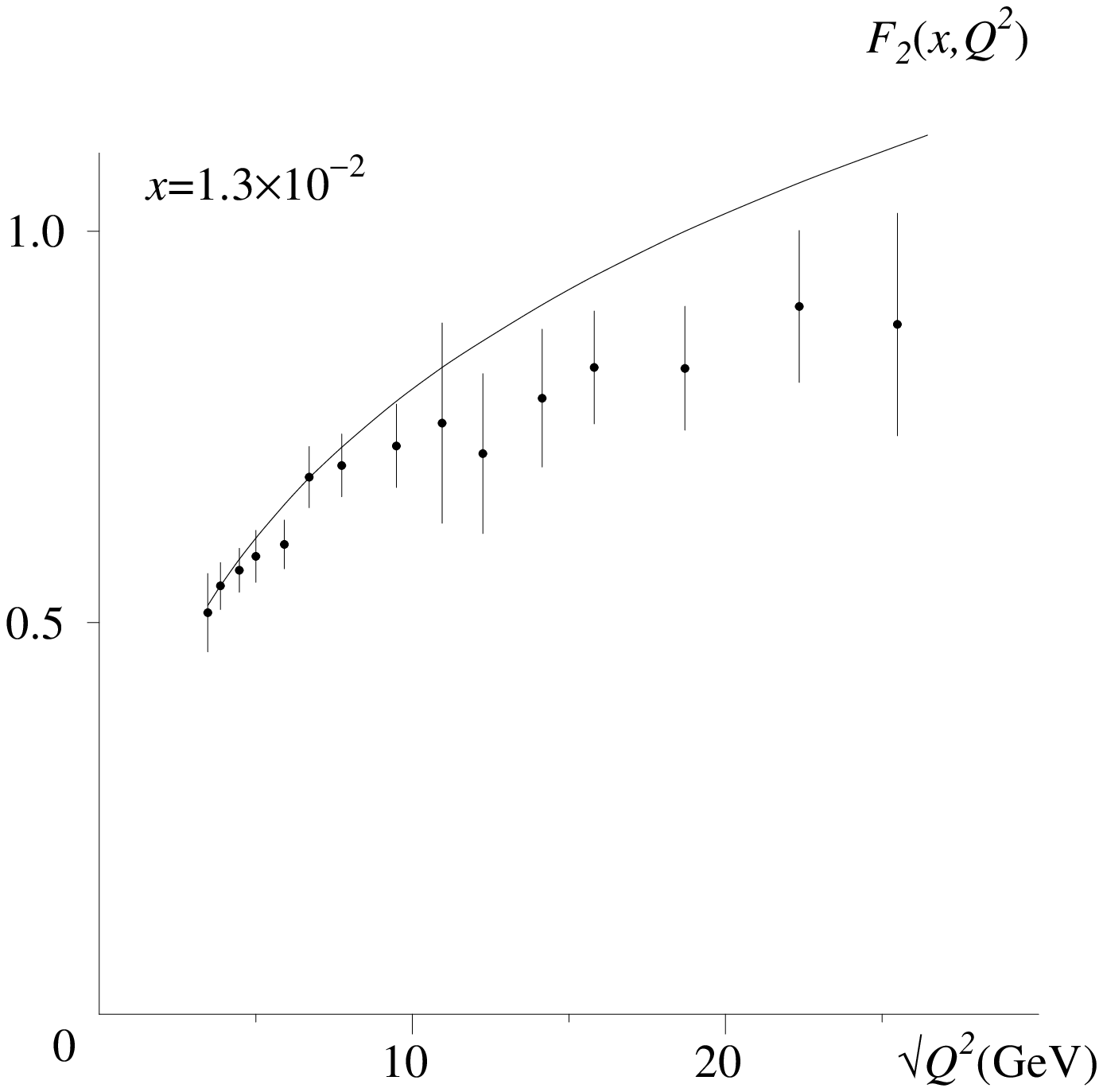}}
\setbox2=\vbox{\hsize75mm\epsfxsize=75truemm\epsffile{ 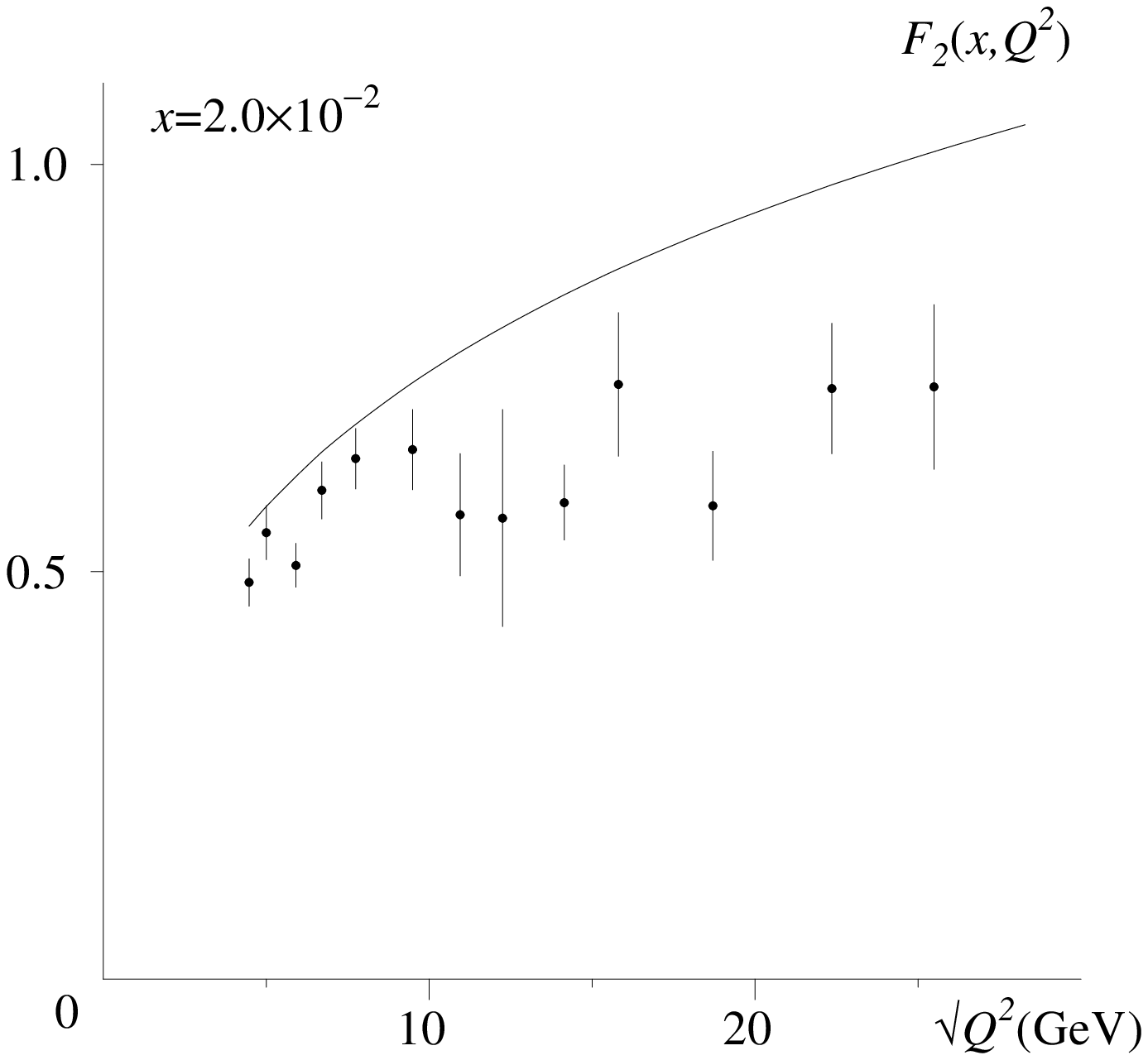}}
\line{\box1\hfill\box2}
\vskip.1cm
\setbox1=\vbox{\hsize78mm{\bf Figure} 6a.- Comparison of theory and \hb\phantom{x}\kern2em
experiment for $x=0.013$.}
\setbox2=\vbox{\hsize78mm{\bf Figure} 6b.- Comparison of theory and \hb\phantom{x}\kern2em 
experiment for $x=0.02$.}
\line{\box1\kern3.em\box2}
\vskip.3cm
$\Deltav_j$ is given in terms of $\hat{B}_{Sj},\,\lambda_j$ by a formula like (5.7b).
 The $\hat{B}_{Sj},\,\lambda_j$ are parameters,
 not given by the present analysis beyond the condition $B_{Si}>0$ which 
follows from positivity of  $F_2$. Choosing 
them appropriately it would not be difficult to obtain that,
 although for every fixed $Q^2$ only 
one term dominates, the envelope of (4.11) when both
 $Q^2\rightarrow \infty,\,x\rightarrow 0$ is (5.2).

In order to discriminate between the two possibilities, we note that, when, 
 for a fixed $x$, the subleading terms become sizeable, they are {\it negative}, 
if they are like in eq. (5.7); while if they are 
as in (4.11) they will start by giving a {\it positive} contribution (until, 
in the end, the term $\Deltav$ also here starts being important). We may thus 
 employ the following strategy. We fit data points including slightly 
larger values of $x$ than before. To be precise, we choose to add to the fit the 
values of $F_2(x,Q^2)$ with $x=1.3\times 10^{-2}$ and still we take
$$12\;{\rm GeV}^2\leq Q^2 \leq 350\;{\rm GeV}^2.$$
 We keep $\lambda=0.38,\Lambdav=0.10\;{\rm GeV}$ fixed. We get a reasonably good fit, 
with $\chi^2/$d.o.f.$=85.1/(73-2)$, $B_S=2.14\times10^{-3},\;B_{NS}=0.54$. 
The comparison of the result, for $x=1.3\times10^{-2}$ and experiment is shown in Fig.5a 
where we have added two points with $Q^2=500\;{\rm and}\;650\;{\rm GeV}^2$ not 
included in the fit; and, for $x=2.0\times10^{-2}$ in Fig. 5b, whose experimental
 points, however, were also {\it not} included in the fit.
 
 Although the results are not conclusive (one should 
have had more care in the treatment of other subleading effects, in 
particular those mentioned in the previous section) it would appear that 
data lie {\it below} theory, thus suggesting (5.7)\footnote*{In fact 
one may fit the difference between the experimental points and the
 dominating contribution by a formula like (5.7b),
 although one should not attach too much 
meanig to this.}  and disfavouring the 
BFKL conjecture, at least at the values of $x,\,Q^2$ attained at HERA.
\vfill
\eject
\noindent 6.-{\bf PROOF OF EQS.}(2.8), (5.1){\bf , SUBLEADING 
CORRECTIONS AND DISCUSSION OF ALTERNATIVES}
\vskip.3cm

We repeat here the proof of the asymptotic behaviour  (2.8), (5.1) for 
ease of reference and with a view to the study of subleading corrections, and 
to discuss the alternatives.
Our assumption is that at a certain, fixed  $Q^2_0$ sufficiently large 
for perturbation theory to be valid we have
$$F_i(x,Q_0^2)=f_i(Q_0^2)x^{-\lambda_i(Q_0^2)}+O(x^{-\mu_i}),\;\mu<\lambda_i(Q_0^2)
,\eqno (6.1)$$
for $i=S,\,G$. Taking moments with {\it continuous} index $n$ as in eq.(2.5) we find 
that the first singularity of the $\mu_i(n,Q^2_0)$ as functions of 
$n$ occur when the integrals diverge, i.e., for $n=1+\lambda_i(Q_0^2)$. {\it Conversely}, 
if for any other $Q^2$ we have moments diverging, 
$$\mu_i(n,Q^2)\sim \frac{1}{n-\lambda_i-1}$$
then necessarily one must have 
$$F_i(x,Q^2)\simeqsub_{x\rightarrow 0}f_i(Q^2)x^{-\lambda_i}.\eqno (6.2)$$
Because of the evolution equations for the 
moments,
$$\left(\matrix{\mu_S(n,Q^2)\cr \mu_G(n,Q^2)\cr}\right)=
\left[\frac{\alpha_s(Q_0^2)}{\alpha_s(Q^2)}\right]^{{\bf D}(n)}
\left(\matrix{\mu_S(n,Q_0^2)\cr \mu_G(n,Q_0^2)\cr}\right),\eqno (6.3)$$
the singularities of the $\mu_i(n,Q^2)$ must originate either in 
those of $\mu_i(n,Q^2_0)$ or in those of the anomalous dimension matrix ${\bf D}(n)$. 
By inspection, the last is seen to be analytic to the right of $n=1$: so 
the {\it leading} 
singularity of the $\mu_i(n,Q^2)$ is identical to that of $\mu_i(n,Q^2_0)$. Thus we 
may write   
$$F_i(x,Q^2)\simeqsub_{x\rightarrow 0}f_i(Q^2)x^{-\lambda_i} \eqno (6.4)$$
with $\lambda_i\equiv\lambda_i(Q_0^2)$
and thus independent of $Q^2$. The rest of the formulas (2.8) (say) follow 
by diagonalizing the matrix ${\bf D}(n)$ and thus reducing the case to 
the evolution of a combination of the structure functions: actually, the 
coefficient relating $F_S$ and $F_G$ in (2.8f) is just 
the corresponding element of this diagonalizing matrix.

Let us next turn to the corrections. Subtracting from e.g. $F_S$ the leading 
singularity we get
$$F_S(x,Q^2)=B_S[\alpha_s(Q^2)]^{d_+(1+\lambda)}\,x^{-\lambda}+F_S^{(1)}(x,Q^2).$$
We can repeat the analysis for $F_S^{(1)}$ provided it behaves, at a fixed 
$Q_1^2$, as $\sim x^{-\lambda^{(1)}}$, $\lambda^{(1)}>0$. In this way we get a series 
like that in (5.9). If, however, the first singularity 
of $\mu_i(n,Q^2_0)$ (once removed that at $n=1+\lambda$) lies to the left 
of $n=1$, then the dominating singularity is that of ${\bf D}(n)$, and we 
get the 
result reported in (5.7). 

From the preceding analysis it follows that 
$\lambda, \,B_S$ cannot have a {\it perturbative}
 dependence on $Q^2$; but they may, and likely 
will, have an indirect dependence via a dependence on the number of flavours excited 
at a given $Q^2$. This has been shown to occur by Witten$^{[19]}$ in the case 
of DIS on {\it photon} targets, where one can {\it calculate} $\lambda$,and  one 
 finds a slight dependence of $\lambda$ on 
$n_f$. This dependence (quite generally now) comes about 
as follows. As discussed before, the behaviour of the structure functions 
is related to the singularities of the moments at  a given $Q^2$. Because 
of the operator product expansion, one has
$$\mu(n,Q^2_0)=C_n\langle p|O_n(Q^2_0)|p\rangle,\eqno (6.5{\rm a})$$
and
$$\mu(n,Q^2)=
C_n\left[\frac{\alpha_s(Q_0^2)}{\alpha_s(Q^2)}\right]^{{\bf D}(n)}
\langle p|O_n(Q^2_0)|p\rangle,\eqno (6.5{\rm b})$$
where $C_n$ are the Wilson coefficents, and the  
$O_n(Q^2_0)$ combinations of quark and gluon operators, renormalized 
at $Q^2_0$, that we rather sketchily write as
$$O_n\sim G\undersetbrace{n}\to{\partial\dots\partial} G;
\;\bar{q}_f\undersetbrace{n}\to {\partial\dots\partial} q_f,$$
and $f=u,d,s,c\dots$ runs over the varios flavours excited. By inspection, the 
singularities of $C_n,\,{\bf D}(n)$ are seen to start at $n=1$, 
both at leading and next to leading order. [For the 
singlet case; for the nonsinglet the 
relevant quantity is $D_{11}(n)$ whose first singularity lies at $n=0$. We send to 
refs. 5,6 for the details of study of this case]. Therefore 
the leading singularities are those of the operator matrix 
elements $\langle p|O_n|p\rangle$, depending on the number of quark flavours that 
intervene. Because the more operators intervene, 
the more poles one is likely to get, this analysis also shows that one 
must have
$$\lambda(n_f=5)\geq\lambda(n_f=4)\geq\lambda(n_f=3)\geq\lambda(n_f=2),$$
an expectation confirmed by our findings in this work.

This argument also shows that the number of flavours to be consider 
should depend mostly on $Q^2$ (and not on the energy, $\nu$). In fact, 
the integral that gives the moments in terms of the 
structure function recives contributions 
from all $x$, and not only from $x\sim 0$.

To finish this section we will discuss what happens if the 
assumption (6.1) fails. That is to say, if at all $Q^2\leq Q^2_0$ with 
$Q^2_0$ a momentum at which one may join the nonperturbative Pomeron 
regime, and perturbative QCD (assumed to exist) one has
$$F_i(x,Q^2_0)\sim O(x^0).$$
This was discussed long ago by De R\'ujula et al.$^{20}$ and has been 
revived recently (see e.g. ref. 21). As explained above, this means that the 
singularities in $n$ of the $\langle p|O_n(Q^2_0)|p\rangle$ lie at, or to the left of 
$n=1$ [else we would have had (6.1)]. Then the 
behaviour of the $F_i$ will be dictated by the 
singularities of ${\bf D}(n)$, for $i=S,\,G$ or $D_{11}(n)$ for 
the nonsinglet case. The analysis is essentially like for 
the corrections $\Deltav$ to $F_2$ discussed in the previous 
section and we immediately obtain,
$$F_2(x,Q^2)\sim\exp \sqrt{b(\log(\log Q^2/\log Q_0^2))|\log x|},
\eqno (6.6{\rm a})$$
and, for the nonsinglet,
$$F_{NS}(x,Q^2)\sim
\exp \sqrt{b'(\log(\log Q^2/\log Q_0^2))|\log x|}.
\eqno (6.6{\rm b})$$
The constants $b,\,b'$, and the more detailed form of 
the behaviour depend on the assumption one makes at $Q^2=Q^2_0$; see e.g. ref. 21.
 For example, if we assume that $F_S(x,Q^2_0)\sim C$, then the formula 
for $F_S(x,Q^2_0)$ becomes like (5.7b), with the obvious changes and in particular,
$$b=\frac{144}{33-2n_f}.$$

It is perhaps not idle to emphasize that (6.6) and (6.1) are mutually {\it exclusive}. 
If, for any $Q^2$ one has (6.1) with $\lambda$ strictly positive,
 one necessarily gets (2.8) and, conversely. Only 
experiment may discriminate between the two alternatives since the 
value of $\lambda$ is not fixed by perturbative QCD. Thus
 either (2.8) or (6.6), but not both, will 
fit experiment if the last is sufficiently precise. 

The fits with (2.8) have been described in the previous sections. In what regards 
eqs.(6.6), I have found it 
 impossible to fit {\it actual} experimental data with them. Eyeball inspection shows that 
 {\it all}the NS structure functions behave as $x^{\frac{1}{2}}$ near $x=0$; 
and evey nonbiased fit, including those in refs.10, 11 (not to 
mention ours in the present paper!) give an $F_2$ behaving also as a power  
$x^{-\lambda}$, with similar exponents $\lambda$
 for all $Q^2$ (between $0.23$ and $0.40$) in the region 
of HERA data. On the other hand, 
any fit with eqs. (6.6), performed fixing $x$ to get rid of the 
large uncertainties in (6.6), give impossibly large values of 
$F_2$ as $Q^2$ grows beyond $100\, \gev^2$, or impossibly small for $Q^2\lsim 40\,\gev^2$.

\vfill\eject
\noindent{\bf REFERENCES}
\vskip.3cm
\noindent 1.- F. J. Yndur\'ain, {\sl Quantum Chromodynamics}, Springer 1983; second 
edition as {\sl The Theory of Quark and Gluon Interactions}, Springer 1992.\hb
2.- V. N. Gribov and L. N. Lipatov, Sov. J. Nucl. Phys. {\bf 15} (1972) 438; 
L. N. Lipatov, {\it ibid.} {\bf 20} (1975) 95:
 Yu. L. Dokshitzer, Sov. Phys. JETP {\bf 46} (1977) 641.\hb
3.- G. Altarelli and G. Parisi, Nucl. Phys. {\bf B126} (1977) 298.\hb
4.- D. G. Gross and F. Wilczek, Phys. Rev. {\bf D9} (1974) 980; 
H. Georgi and H. D. Politzer, {\it ibid}, 416.\hb
5.- C. L\'opez and F. J. Yndur\'ain, Nucl. Phys. {\bf B171} (1980) 231.\hb
6.- C. L\'opez and F. J. Yndur\'ain, Nucl. Phys. {\bf B183} (1981) 157.\hb
7.- A. Gonz\'alez-Arroyo, C. L\'opez and F. J. Yndur\'ain, Phys. Lett. 
{\bf 98B} (1981) 215.\hb
8.- B. Escoubes et al., Nucl. Phys. 
{\bf B242} (1984) 329.\hb
9.- K. Prytz, Phys. Lett. {\bf B332} (1994) 393.\hb
10.- R. K. Ellis, Z. Kunszt and E. M. Levin, Nucl. Phys. {\bf B420} (1994) 517.\hb
11.-M. Derrick et al, Z. Phys. {\bf C65} (1995) 397; Phys. Lett. {\bf B345} 
(1995) 576.\hb
12.-C. L\'opez and F. J. Yndur\'ain, Phys. Rev. Lett. {\bf 44} (1980), 1118.\hb
13.-T. H. Bauer, R. D. Spital and D. R. Yennie, Rev. Mod. Phys. {\bf 50} (1978) 261 
for a review and references.\hb
14.-F. J. Yndur\'ain, Plys. Lett. {\bf 41B} (1972) 591.\hb
15.-Particle Data Group, Phys. Rev.\hb
16.- M. Derrick et al, Phys Lett. {\bf B293} (1992) 465; Z. Phys. 
{\bf C63} (1995) 391.\hb
17.- Yu. A. Simonov,Yadernaya Fizika, {\bf 58} (1995) 113, and work quoted there.\hb
18.- E. A. Kuraev, L. N. Lipatov and V. S. Fadin, Sov. Phys. JETP {\bf 44} (1976) 443; 
Ya. Ya. Balitskii and L. N. Lipatov, Sov. J. Nucl. Phys. {\bf 28} (1978) 822.\hb
19.- E. Witten, Nucl. Phys. {\bf B120} (1977) 189.\hb
20.- A. De R\'ujula et al., Phys. Rev{\bf D10} (1974) 1649.\hb
21.- S. Forte and R. D. Ball, CERN -TH/ 95-323 (1995).
\vskip2cm
\noindent{\bf ACKNOWLEDGEMENTS}
\vskip.3cm
The author is grateful to F. Barreiro (who had performed a preliminary fit to the old HERA data with our 
formulas) for communicating his results, for discussions and for information on the as yet 
unpublished more recent data. Thanks are also due to K. Adel, whose program ``Kdraw" has 
been used for the figures.

The partial financial support of CICYT, Spain, is also acknowledged.

\bye